\documentclass[prl,twocolumn,tightenlines,letterpaper,showpacs,groupedaddress,superscriptaddress,nofootinbib,floatfix,preprintnumbers,tightenlines,10pt]{revtex4-1}
\usepackage[bookmarks=false]{hyperref}
\usepackage{amssymb,amsmath,graphicx,color}
\usepackage{bm}
\usepackage[tight]{subfigure}

\def\si{^1 \hskip -0.03in S _0}
\def\siii{^3 \hskip -0.025in S _1}

\def\pislash{{\pi\hskip-0.55em /}}

\def\L1Abar{\tilde{l}_{1,A}}

\def\L{{\Lambda}}

\def\cO{{\mathcal O}}

\def\cR{{\mathcal R}}

\def\cW{{\mathcal W}}

\def\tnubb{$2\nu\beta\beta$}
\def\znubb{$0\nu\beta\beta$}
\def\eqref#1{{(\ref{#1})}}

\def\Red{}

\begin{document}

\begin{figure}[!t]
\vskip -0.5cm
{
\includegraphics[width=3.0 cm]{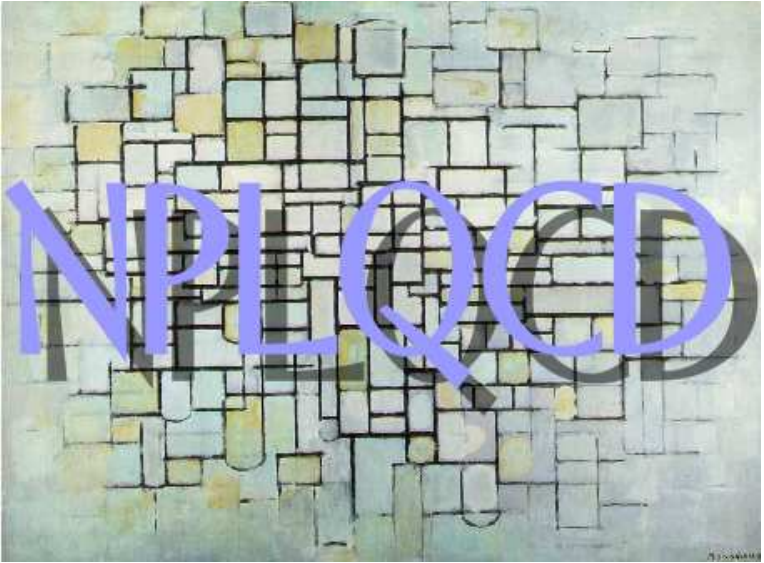}
}
\vskip -0.5cm
\end{figure}

\title{
Isotensor axial polarisability and lattice QCD input for \\ nuclear  double-$\beta$ decay phenomenology 
}

  \author{Phiala E. Shanahan } \affiliation{
  	Center for Theoretical Physics, 
  	Massachusetts Institute of Technology, 
  	Cambridge, MA 02139, USA}

 \author{Brian C. Tiburzi} 
\affiliation{ Department of Physics, The City College of New York, New York, NY 10031, USA }
\affiliation{Graduate School and University Center, The City University of New York, New York, NY 10016, USA }

\author{Michael L. Wagman} 
\affiliation{Department of Physics,
	University of Washington, Box 351560, Seattle, WA 98195, USA}
\affiliation{Institute for Nuclear Theory, University of Washington, Seattle, WA 98195-1550, USA}

\author{Frank Winter}
\affiliation{Jefferson Laboratory, 12000 Jefferson Avenue, 
	Newport News, VA 23606, USA}

\author{Emmanuel~Chang}
\affiliation{Institute for Nuclear Theory, University of Washington, Seattle, WA 98195-1550, USA}

 \author{Zohreh Davoudi} \affiliation{
 	Center for Theoretical Physics, 
 	Massachusetts Institute of Technology, 
 	Cambridge, MA 02139, USA}

 \author{William Detmold} \affiliation{
 	Center for Theoretical Physics, 
 	Massachusetts Institute of Technology, 
 	Cambridge, MA 02139, USA}

 \author{Kostas~Orginos}
 \affiliation{Department of Physics, College of William and Mary, Williamsburg,
 	VA 23187-8795, USA}
 \affiliation{Jefferson Laboratory, 12000 Jefferson Avenue, 
 	Newport News, VA 23606, USA}
 
 \author{Martin J. Savage}
 \affiliation{Institute for Nuclear Theory, University of Washington, Seattle, WA 98195-1550, USA}

\collaboration{NPLQCD Collaboration}

\date{\today}

\preprint{INT-PUB-16-056, MIT-CTP-4867}


\begin{abstract}
The potential importance of short-distance nuclear effects in double-$\beta$ decay is assessed using a lattice QCD calculation of the $nn\rightarrow pp$ transition and effective field theory methods. 
At the unphysical quark masses used in the numerical computation, these effects, encoded in the isotensor axial polarisability, are found to be of similar magnitude to the nuclear modification of the single axial current, which phenomenologically is the quenching of the axial charge used in nuclear many-body calculations. 
This finding suggests that nuclear models for neutrinoful and neutrinoless double-$\beta$ decays should incorporate this previously neglected contribution if they are to provide reliable guidance for next-generation neutrinoless double-$\beta$ decay searches. The prospects of constraining the isotensor axial polarisabilities of nuclei using lattice QCD input into nuclear many-body calculations are discussed.
\end{abstract}

\maketitle
Double-$\beta$ ($\beta\beta$) decays of nuclei are of significant phenomenological interest; they probe fundamental symmetries of nature and admit both tests of the Standard Model (SM) and investigations of physics beyond it \cite{Mohapatra:2005wg}. Consequently, these decays are the subject of intense experimental study, and next-generation  $\beta\beta$-decay experiments are currently being planned \cite{Avignone:2007fu,Bilenky:2012qi,Dell'Oro:2016dbc}.
At present, both the robust prediction of the efficacy of different detector materials, necessary for optimal design sensitivity, and the robust interpretation of the highly sought-after neutrinoless $\beta\beta$-decay (\znubb) mode are impeded by the lack of knowledge of second-order weak-interaction nuclear matrix elements. These quantities bear uncertainties from nuclear modelling that are both significant and difficult to quantify  \cite{Engel:2016xgb}. 
Controlling the nuclear uncertainties in $\beta\beta$-decay matrix elements by connecting the nuclear many-body methods to the underlying parameters of the SM is a critical task for nuclear theory.

In this Letter, lattice QCD and pionless effective field theory (EFT($\pislash$)) are used to investigate the strong-interaction uncertainties in the second-order weak transition of the two-nucleon system in the SM by determining the
threshold transition matrix element for $nn\to pp$. This matrix element receives long-distance contributions from the deuteron intermediate state whose size is governed by the squared magnitude of the $\langle pp | \tilde{J}_\mu^+ | d \rangle$ matrix element of the axial current that has been recently calculated using lattice quantum chromodynamics (LQCD)~\cite{Savage:2016kon}. In that work, the two-body contribution to the matrix element ({\it i.e.}, that beyond the coupling of the axial current to a single nucleon) was constrained, quantifying the effective modification ({\it quenching}) of the axial charge of the nucleon from two-body effects.
Here, it is highlighted that the $nn\to pp$ matrix element receives additional short-distance contributions beyond those in $|\langle pp | \tilde{J}_\mu^+ | d \rangle|^2$ arising from the two axial currents being  separated by $r<\Lambda^{-1}\sim m_\pi^{-1}$ (where $\Lambda$ is the cutoff scale of EFT($\pislash$)), referred to herein as the {\it isotensor axial polarisability}. Using EFT($\pislash$) to analyse both the second-order weak transition calculated here for the first time and the first-order $\langle pp | \tilde{J}_\mu^+ | d \rangle$ transition~\cite{Savage:2016kon}, the short- and long-distance contributions to the $nn\to pp$ matrix element are separately determined. Interestingly, the short-distance contribution to the total matrix element from the axial polarisability is found to be 
of comparable size  (within the uncertainties of the present calculation) to the 
two-nucleon current contribution to $|\langle pp | \tilde{J}_\mu^+ | d \rangle|^2$. Described in phenomenological terms, the polarisability thus appears to be as important as the effective quenching of the axial charge in the two-nucleon system. 

The numerical calculations in this work are performed at unphysical values of the quark masses and for a disallowed decay. While there is no immediate phenomenological impact of the numerical values of the matrix elements that are extracted, the observed behavior does provide an important lesson for many-body calculations.
In typical calculations of two-neutrino (\tnubb) decay, the nuclear matrix elements are calculated using two insertions of the axial current in a truncated model space, with a quenched value of $g_A$ tuned to reproduce experiment. 
If the findings presented here persist at the physical values of the quark masses,  they would imply that a 
potentially significant contribution has been ignored in standard \tnubb\ calculations, resulting in a source of uncertainty in the nuclear matrix elements that remains to be quantified. 
Importantly, this uncertainty can only be constrained  using $\beta\beta$-decay measurements or numerical calculations. In \znubb\ decays, the situation becomes even less certain, in part due to dependence on possible scenarios of physics beyond the SM. With a light Majorana neutrino, generalisations of the axial polarisability will also likely be relevant.

In what follows, the lattice QCD and EFT($\pislash$) calculations and the analysis of the axial polarisability are summarised, with complete details presented in a subsequent paper \cite{Tiburzi:2017iux}. The potential for future lattice QCD calculations to provide the necessary input to constrain many-body calculations of \tnubb\ and \znubb\ matrix elements, and thereby reduce the uncertainties in calculated $\beta\beta$-decay rates, is also discussed.

\vspace*{3mm}
\noindent{\it Two-neutrino $\beta\beta$-Decay:} 
The focus of this Letter is on \tnubb\ decay of the dinucleon system. The decay width is given by 
\begin{eqnarray}
[T^{2\nu}_{1/2}]^{-1}&=&
G_{2\nu}(Q) |  M_{GT}^{2\nu}|^2,
\label{eq:DW}
\end{eqnarray}
where $Q=E_{nn}-E_{pp}$,  $G_{2\nu}(Q)$ is a known phase-space factor \cite{Kotila:2012zza,Stoica:2013lka}, and the Gamow-Teller matrix element in the two-nucleon system is
\begin{eqnarray}
M_{GT}^{2\nu}&=& 
6 \times \frac{1}{2}
\int{d^4x} \, {d^4y} 
\, 
\langle pp |  T \left[ J^+_{3}(x) J^+_{3}(y)  \right] | nn \rangle 
\nonumber
\\& =& 6
 \sum_{{\frak l'}}\frac{\langle pp |  \tilde{J}_3^+ |{\frak l'}\rangle\langle {\frak l'} |  \tilde{J}_3^+ |nn \rangle}{E_{\frak l'}-(E_{nn}+E_{pp})/2}.
\label{eq:MGT}
\end{eqnarray}
Here, $J_3^+=(J_3^1+i J_3^2)/\sqrt{2}$ is the  $3$rd-component of the  $\Delta I_3=1$ axial-vector current, 
$J_\mu^a(x) = \overline {q}(x) \gamma_\mu\gamma_5 \frac{\tau^a}{2} q(x)$, and ${\frak l'}$ indexes a complete set of zero-momentum hadronic states with the quantum numbers of the deuteron. The factors of 6 in Eq.~(\ref{eq:MGT}) are due to rotational symmetry and our normalization of the currents. 
We employ 
$\tilde{J}^+_3 = \int d\bm{x} \, J_3^+ (\bm{x}, t=0)$
to denote the zero-momentum current at 
$t = 0$.

As with forward Compton scattering, the amplitude can be written in terms of a Born term, corresponding to an intermediate deuteron state, and the isotensor axial polarisability which absorbs the contributions from the remaining states in the above summation. By isospin symmetry, this polarisability is most cleanly identified  as the forward  matrix element of the $I=2$, $I_3=0$ component of the time-ordered product of two axial-vector currents in the $\si$ $np$ ground-state with the deuteron pole (the Born term) omitted.  
For use below, isospin relations allow  this matrix element to be written as
\begin{eqnarray}
\langle p p | J^+_{3}(x) J^+_{3}(y) | n n \rangle 
&=&
\langle n p | J_3^{(u)}(x) J_3^{(u)}(y) | n p \rangle 
\nonumber \\
&&
-\frac{1}{2}  \langle n n | J_3^{(u)}(x) J_3^{(u)}(y) | n n \rangle 
\nonumber \\
&&
-\frac{1}{2}  \langle n n | J_3^{(d)}(x)  J_3^{(d)}(y) | n n \rangle 
\label{eq:recipe},
\quad
\end{eqnarray}
where $J_3^{(q)}(x) = \overline{q}(x) \gamma_3\gamma_5 q(x)$.

\vspace*{3mm}
\noindent{\it Pionless effective field theory:}
EFT($\pislash$)~\cite{Kaplan:1996nv,Kaplan:1998tg, Kaplan:1998we, vanKolck:1998bw, Chen:1999tn, Beane:2000fi}  efficiently describes two-nucleon systems in the regime where momenta are small compared to the pion mass. This is an appropriate tool with which to address \tnubb\ decays at heavier quark masses, but the inclusion of explicit pion degrees of freedom will likely be required at the physical quark masses (\znubb\ decay probes higher momenta, $k\sim 100$ MeV, in large nuclei and likely also requires an EFT with  explicit pion degrees of freedom).
In what follows, the dibaryon formalism of EFT($\pislash$) is utilised, using the conventions for the strong-interaction sector described in Ref. \cite{Beane:2000fi}. The nucleon degrees of freedom are encoded in the field $N$, and the two-nucleon degrees of freedom enter as the  isosinglet, $t_i$, and isotriplet, $s_a$, dibaryon fields while  $y_t$ and $y_s$ describe the couplings between two nucleons and the corresponding dibaryon fields.
In this formalism, the single axial-current interactions enter through the Lagrangian \cite{Butler:1999sv, Butler:2000zp, Kong:2000px,Butler:2001jj}
\begin{eqnarray}
{\cal L}^{(1)} & = & -\frac{g_A}{2} N^\dagger W_3^a \sigma_3 \tau^a N
\nonumber \\
&& +\left(g_A-\frac{\tilde{l}_{1,A}} {2M\sqrt{r_s r_t}} \right) \left(W_3^a t_3^\dagger s^a +\text{h.c.} \right)
,
\label{eq:L-dibaryon-1}
\end{eqnarray}
 where $r_{s(t)}$ is the effective range in the $\si(\siii)$ two-nucleon channel, $\sigma_i(\tau^a)$ are Pauli matrices in spin(flavour) space, $g_A$ and $\tilde l_{1,A}$ are the one- and two-nucleon axial couplings, and  $W_3^a$ is an  axial isovector field aligned in the $j=3$ spatial direction. The second term is constructed so that $\tilde l_{1,A}$ corresponds to a purely two-body current effect.
The second-order isotensor axial interaction in the $\si$ channel enters as
\begin{eqnarray}
{\cal L}^{(2)} & = & - \left(\frac{Mg_A^2}{4\gamma_s^2}+\frac{\tilde{h}_{2,S}} {2Mr_s}\right)  {\cW}^{ab} {s^a}^\dagger s^b,
\label{eq:L-dibaryon-2}
\end{eqnarray}
where $\cW^{ab}=W_3^{\{a}W_3^{b\}}$ is the traceless symmetric combination of two axial fields at the same location, $\tilde h_{2,S}$ is the scalar isotensor weak two-nucleon coupling and $\gamma_{s}=\sqrt{M B_{nn}}$  with  the binding energy  of the $\si$ system being $B_{nn}$ (at the unphysical masses used herein, the $\si$ system is bound \cite{Beane:2012vq}).

Calculation of the $nn\to pp$ amplitude, presented in detail in Ref.~\cite{Tiburzi:2017iux}, shows that 
\begin{eqnarray}
\label{eq:M-nnpp-EFTb}
M_{nn \to pp}
& = & 
-\frac{|\langle pp | \tilde{J}_3^+ |d \rangle|^2}{\Delta} 
\ +\ 
\frac{M g_A^2}{4 \gamma_s^{2}} 
-\mathbb{H}_{2,S}
\end{eqnarray}
where  $\Delta=E_{nn}-E_d$ is the difference of the ground-state $\si$ and $\siii$ energies. The parameter
$\mathbb{H}_{2,S}$ is the short-distance two-nucleon, two-axial current coupling $\tilde{h}_{2,S}$ of Eq.~(\ref{eq:L-dibaryon-2}), redefined and rescaled to capture the effects beyond the deuteron pole and  two-nucleon states at energies below $\Lambda$ \cite{Tiburzi:2017iux}. 
This expression depends on both the long-distance contribution from the deuteron pole (the first term) and the short-distance contributions encapsulated in the second and third terms (where ``short-distance'' here means the ``nonadiabatic'' contribution; that is, every process other than those proceeding via the bound $\siii$ ground state).
The deuteron-pole contribution includes the effective quenching of the axial charge through $\tilde l_{1,A}$.
A determination of the $nn\to pp$ transition matrix element, along with an extraction of  the $pp \to d $ amplitude,  
allows for the isolation of the unknown short-distance contribution, $\mathbb{H}_{2,S}$. 
Once this counterterm is determined, few-body methods based on EFT($\pislash$) or matched to them (for example, see Ref.~\cite{Barnea:2013uqa}) can incorporate the axial polarisability in computations of decay rates of larger nuclei.

\vspace*{3mm}
\noindent{\it Lattice QCD calculations:} The present lattice QCD calculations extend those of the $pp$-fusion cross section and tritium $\beta$-decay in Ref.~\cite{Savage:2016kon}. The same hadronic correlators calculated in the presence of  external axial fields are analysed further to access second-order weak responses to the external field. 
Recent calculations by the RBC/UKQCD collaboration in the kaon sector~\cite{Christ:2012se,Christ:2015pwa,Christ:2016eae,Christ:2016mmq} have demonstrated that long-distance second-order weak effects can be constrained using lattice QCD. These methods are extended to determine the second-order weak matrix elements of the two-nucleon system.

As discussed in Ref.~\cite{Savage:2016kon}, calculations are performed on one ensemble of gauge-field configurations generated using a L\"uscher-Weisz gauge action~\cite{Luscher:1984xn} and a clover-improved fermion action~\cite{Sheikholeslami:1985ij} with $N_f = 3$ degenerate flavours of quarks. 
The quark masses are tuned to the physical strange-quark mass, producing a pion of mass $m_\pi \approx 806~{\rm MeV}$.  The ensemble has a spacetime volume of $L^3\times T=32^3\times48$ and a gauge coupling that corresponds to a lattice spacing of $a\sim 0.145~{\rm fm}$. For these calculations, 437 configurations spaced by 10 hybrid Monte Carlo trajectories are used and seven  different sets of compound propagators are generated from sixteen smeared sources on each configuration with both smeared (SS) and point (SP) sinks. The compound propagators are produced with a single insertion of $J^{(u,d)}_3$ with couplings $\lambda_{u,d}= \{0, \pm 0.05, \pm 0.1, \pm 0.2\}$ . These propagators are used to produce correlation functions 
\begin{eqnarray}
C^{(h)}_{\lambda_u;\lambda_d}(t) 
& = & 
\sum_{\bf x}
\langle 0| \chi_h({\bf x},t) \chi^\dagger_h(0) |0 \rangle_{\lambda_u;\lambda_d},
\label{eq:bfcorr}
\end{eqnarray}
for all the allowed spin states of the one- and two-nucleon systems,
 $h\in\{p$, $np(\siii)$, $nn$, $np(\si)$, $pp\}$.  Results for all source locations  on each configuration are averaged before subsequent analysis. 

The calculations use a lattice axial current with the finite renormalisation factor $Z_A=0.867(43)$~\cite{Savage:2016kon}. Because of the isotensor nature of the bilinear insertions, mixing with other operator structures is highly suppressed.
As the dinucleon and deuteron states are both compact bound states at this value of the quark masses  \cite{Beane:2012vq}, only exponentially small volume effects are anticipated in the extracted matrix element. This will become a more subtle issue for future calculations with quark masses near the physical values, as discussed in Ref.~\cite{Tiburzi:2017iux}.

The second-order axial responses of the dinucleon system are the primary focus of the current work. 
For an up-quark axial current, the relevant background-field correlators have the form 
\begin{eqnarray}
C^{(h)}_{\lambda_u;\lambda_d=0}(t) 
&=&
\sum_{\bf x}
\langle 0| \chi_{h}({\bf x},t) \chi^\dagger_{h}(0) |0 \rangle  
\nonumber \\ 
&&
\hspace*{-1.9cm}+ \lambda_u
\sum_{{\bf x},{\bf y}}\sum_{t_1=0}^t
\langle 0| \chi_{h}({\bf x},t) J_3^{(u)} ({\bf y},t_1)   \chi^\dagger_{h}(0) |0 \rangle 
\nonumber \\ 
&&\hspace*{-1.9cm}+ \frac{\lambda_u^2}{2}
\sum_{{\bf x},{\bf y},{\bf z}}\sum_{t_{1,2}=0}^t 
\langle 0| \chi_{h}({\bf x},t) J_3^{(u)} ({\bf y},t_1)  
J_3^{(u)} ({\bf z},t_2)   \chi^\dagger_{h}(0) |0 \rangle 
\nonumber \\
&&\hspace*{-1.9cm}
+  \cO(\lambda_u^3), 
\label{eq:quad1}
\end{eqnarray}
from which the second-order term in the field strength, $\lambda_u$, can be extracted from determinations at multiple values of $\lambda_u$. Combining these correlators, and those for the down-quark axial current, as specified
in Eq.~(\ref{eq:recipe}), leads to the isotensor matrix element 

\begin{eqnarray}
C(t)&=&\left. 2C^{(np(\si))}_{\lambda_{u};0}(t)\right|_{\lambda_u^2}
-
  \left.C^{(nn)}_{\lambda_{u};0}(t)\right|_{\lambda_u^2}
- 
\left.C^{(nn)}_{0;\lambda_{d}}(t)\right|_{\lambda_d^2},
\nonumber \\
\label{eq:Cnnpp}
\end{eqnarray}
where each term on the right-hand-side is the component of the correlation functions that is second-order in $\lambda_u$ or $\lambda_d$, as denoted. 
Using Eq.~(\ref{eq:recipe}) and isospin symmetry, Eq.~(\ref{eq:Cnnpp}) can be written as
\begin{eqnarray}
C(t)
&=&\sum_{{\bf x},{\bf y},{\bf z}}\sum_{t_{1,2}=0}^t 
\langle 0| \chi_{pp}({\bf x},t) T\left[J_3^{+} ({\bf y},t_1)  
J_3^{+} ({\bf z},t_2) \right]   \chi^\dagger_{nn}(0) |0 \rangle .
\nonumber \\
\end{eqnarray}
Up to discretisation effects, insertion of appropriate complete sets of states allows this expression to
be written as
\begin{eqnarray} 
C(t) & =& 
\frac{2}{a^2} 
\sum_{\frak{n},\frak{m}, \frak{l}'} 
Z_{\frak{n}}Z_{\frak{m}}^\dagger 
e^{- E_\frak{n} t} 
\frac{
\langle \frak{n}| \tilde{J}_3^{+} | \frak{l}' \rangle 
\langle \frak{l}' | \tilde{J}_3^{+}  | \frak{m} \rangle
}{E_{\frak{l}'}- E_\frak{m}} 
\label{eq:CC}
 \\
\nonumber
&&\ \times\left(
\frac{
e^{- \left(E_{\frak{l}'} - E_\frak{n}\right) t} -1
}
{E_{\frak{l}'} - E_\frak{n}}
+
\frac{
e^{\left(E_\frak{n}-E_\frak{m}\right)t}-1
}
{E_\frak{n}-E_\frak{m}}\right)
,
\end{eqnarray}
where 
$|\frak{n} \rangle$, 
$|\frak{m}\rangle$ 
and 
$|\frak{l}' \rangle$  
are zero-momentum energy eigenstates with the quantum numbers of the
$pp$, 
$nn$ 
and deuteron systems, 
respectively. 
Here
$Z_{\frak{n}}= \sqrt{V} \langle 0| \chi_{pp} | \frak{n}\rangle$ 
and
$Z_{\frak{m}} = \sqrt{V} \langle 0| \chi_{nn} | \frak{m}\rangle$ 
are overlap factors, 
and 
$E_{\frak{l}'}=E_{nn}+\delta_{\frak{l}'}$  
and 
$E_\frak{n}=E_{nn}+\delta_\frak{n}$ 
are the energies of the 
$\frak{l}'$th 
and 
$\frak{n}$th 
excited states in the $\siii$ and $\si$ channels, respectively.

Forming a ratio of Eq.~(\ref{eq:CC}) to the zero-field two-point function,
\begin{eqnarray}
\cR(t)&=&\frac{C(t)}{2 C^{(nn)}_{0;0}(t)},
\label{eq:Rnnpp}
\end{eqnarray}
it is straightforward, utilizing the isospin symmetry of the calculation,  
to show that%
~\cite{Tiburzi:2017iux}
\begin{eqnarray}
a^2 
\hat\cR(t)
& = &  
a^2 
\cR(t) - \frac{|\langle pp| \tilde{J}_3^{+}  |d\rangle|^2 }{\Delta} 
\left[ \frac{e^{\Delta t}-1}{\Delta} -t \right] 
\label{eq:rnnppsub}
\\
& \hspace*{-1.7cm}= & 
 \hspace*{-0.7cm} t \sum_{{\frak l}' \ne d}
{
	\langle pp| \tilde{J}_3^{+}  |{\frak l}' \rangle \langle {\frak l}' | \tilde{J}_3^{+}| nn \rangle
	\over E_{{\frak l}'}-E_{nn}}
+ c  + d\ e^{\Delta t} + \cO(e^{-\hat\delta t}) ,
 \nonumber
\end{eqnarray}
where $c$ and $d$ involve complicated combinations of ground- and excited-state transition amplitudes, and $\hat{\delta}$ is the minimum energy gap to the first excited state in either channel; and, for  these calculations, $\hat{\delta} \gg\Delta$. Importantly, the coefficient of the linear term determines the axial polarisability and can be extracted 
from
\begin{eqnarray}
\cR^{\text{(lin)}}(t)= 
{(e^{ a \Delta} +1) \hat\cR(t+a) - \hat\cR(t+2 a) - e^{a \Delta} \hat\cR(t) \over 
 e^{a \Delta} -1 }
\nonumber \\
\label{eq:Rlin}
\end{eqnarray}
at late times.
Finally, this result can be combined with the deuteron-pole contribution to give a quantity that asymptotes to the bare Gamow-Teller matrix element at late times,
\begin{eqnarray}
\cR^{\text{(full)}}(t)= \cR^{\text{(lin)}}(t) - \frac{|\langle pp| \tilde{J}_3^{+}  |d\rangle|^2 }{a \Delta}  \stackrel{t\to\infty}{\longrightarrow} {M^{2\nu}_{GT}\over 6 \, a Z_A^2} . \ \ \
\label{eq:Rfull}
\end{eqnarray}

\begin{figure}[!ht]
	\centering
		\includegraphics[width=\columnwidth]{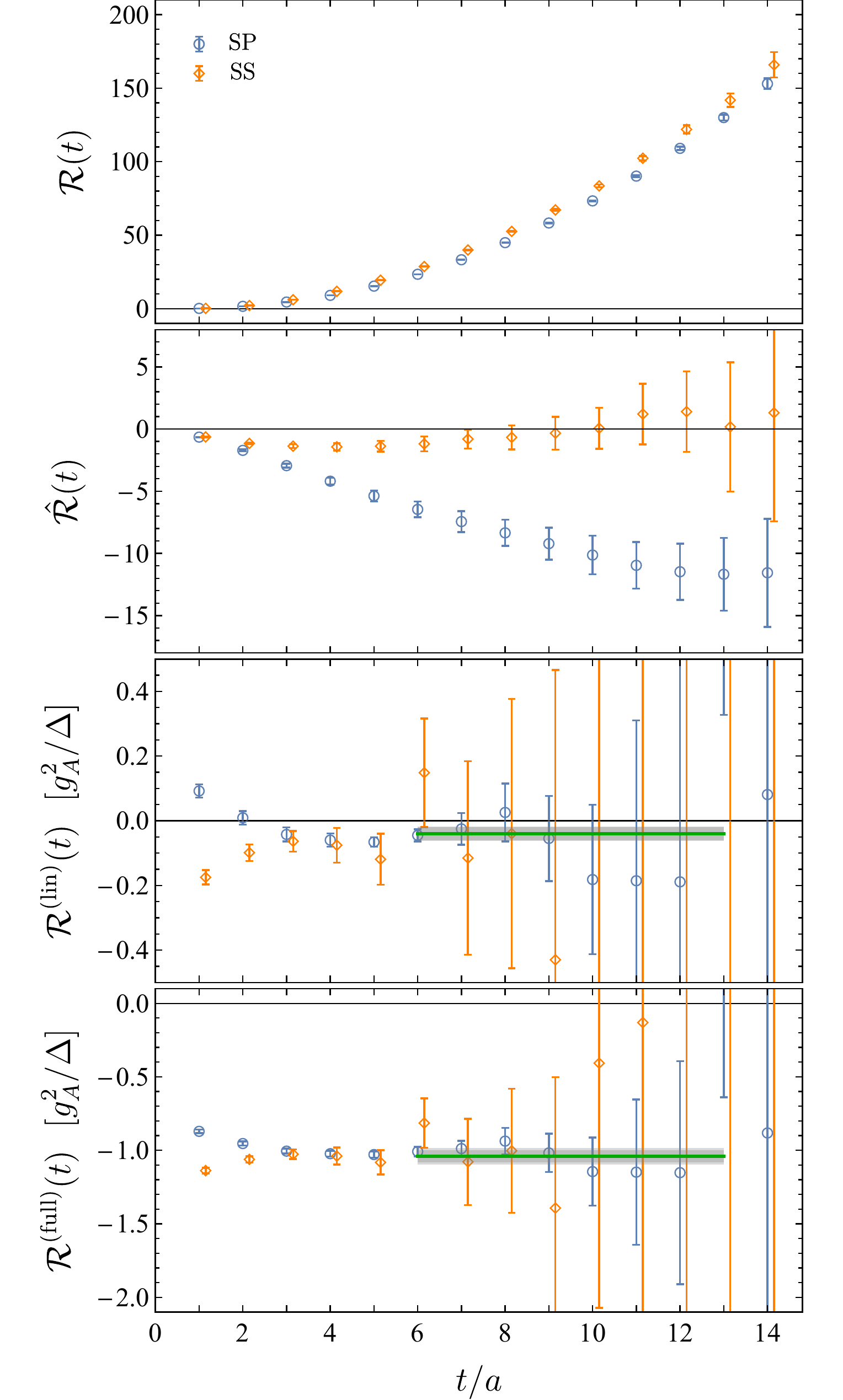}
	\caption{
		Ratios from Eqs.~(\ref{eq:Rnnpp})--(\ref{eq:Rfull}) used in the analysis.
		In each panel, the orange diamonds (blue circles) correspond to the SS (SP) data. The green bands show fits to the SP data in the lower two panels.	The SS data are slightly offset in the horizontal direction for clarity. The difference between the  SS and SP ratios in the upper two panels is due to contamination that is removed in constructing the subsequent quantities in the lower panels.
		}
	\label{fig:MEextract}
\end{figure}

The four ratios used to determine $M_{GT}^{2\nu}$ are shown in 
Fig.~\ref{fig:MEextract} for both SS and SP source--sink combinations.
Fits are performed to the statistically more precise SP correlators and the 
values of the total matrix element and the short-distance contribution, normalised by the naive deuteron-pole matrix element $g_A^2/\Delta$”, are given by
\begin{eqnarray}
{\Delta\over g_A^2}\sum_{{\frak l}' \ne d}
{
	\langle pp| \tilde{J}_3^{+}  |{\frak l}' \rangle \langle {\frak l}' | \tilde{J}_3^{+} | nn \rangle
	\over E_{{\frak l}'}-E_{nn}}
& = & 
\Red{-0.04(2)(1)},
\\{1\over6}
{\Delta\over g_A^2}{M_{GT}^{2\nu} } &=&
\Red{-1.04(4)(4)}
 .
\end{eqnarray}
In these expressions, the first uncertainties arise from statistical sampling and from systematic effects 
from fitting choices and deviations from Wigner symmetry \cite{Tiburzi:2017iux}. The second uncertainties encompass differences between  analysis methods. The leading discretisation effects, which are potentially large on the numerically smaller polarisability term,  are removed by  normalising to the square of the proton axial charge computed using the same lattice axial current on the same ensemble.

\vspace*{3mm}
{\it Discussion:}
The computed value of $M_{GT}^{2\nu}$ above can be used to determine the unknown  EFT($\pislash$) low-energy constant $\mathbb{H}_{2,S}$. Taking  the values of $g_A$ and the two-body single-current matrix element from Ref.~\cite{Savage:2016kon}, and using the calculated binding energies and effective ranges of the two-nucleon systems \cite{Beane:2012vq,Beane:2013br}, the result is \Red{$\mathbb{H}_{2,S} =4.7(1.3)(1.8)\ {\tt fm}$}. 
The dominant contribution to $M_{GT}^{2\nu}$ comes from the deuteron pole with coupling $g_A^2$. This is modified  by two-body effects in the axial current  ($\tilde l_{1,A}$); this contribution shifts the leading-order result from  $g_A^2/\Delta$ to $|\langle pp| \tilde{J}_3^{+}  |d\rangle|^2/\Delta$, which is approximately a \Red{5\% shift}. 
Interestingly, the calculated result suggests that the additional axial polarisability contribution is of similar size.
The existence of this short-distance contribution precludes accurately predicting $\beta\beta$-decay matrix elements in a nuclear many-body calculation by simply rescaling (quenching) $g_A$.

The present results are obtained at an unphysical quark mass without the inclusion of electromagnetism and isospin breaking effects, and at a single lattice spacing and volume. All these caveats may be important,  particularly given that the short-distance two-nucleon effects are only few-percent contributions to the matrix elements. Such effects require further investigation but are not expected to qualitatively alter the conclusions of this work.  Despite these qualifications, 
the potential for a relatively large contribution from the isotensor axial polarisability is important, as terms of this form are not included in current phenomenological analyses of $\beta\beta$-decay. This observation, supported and motivated by the numerical calculations, is the central point of this Letter. In order to accurately 
predict \tnubb\ decay rates with fully quantified uncertainties, the isotensor axial polarisabilities of nuclei must be determined. Future lattice QCD calculations in few-nucleon systems and light nuclei, matched to nuclear many-body methods such as EFT($\pislash$), offer the possibility of determining these contributions which are difficult to access experimentally. 
However, in order to undertake such calculations  at the
physical quark masses, a number of difficulties related to the bi-local nature of such weak processes must be overcome \cite{Tiburzi:2017iux}.
Additional complications arise for \znubb\ where these polarisability contributions are also likely to be important in the case of light Majorana neutrinos.
Furthermore, the short-distance strong interaction contributions encapsulated in the isotensor axial polarisability provide an inherent background to extracting contributions from short-distance lepton-number violating operators that are possible beyond the light Majorana scenario \cite{Savage:1998yh,Prezeau:2003xn,Graesser:2016bpz,Nicholson:2016byl,Cirigliano:2017ymo}. 

\vspace*{3mm}
{\it Acknowledgments:}
This research was supported in part by the National Science Foundation under grant number NSF PHY11-25915 and
ZD, WD, MJS, PES, BCT and MLW acknowledge the Kavli Institute for Theoretical Physics for hospitality 
during completion of this work.
Calculations were performed using computational resources provided
by NERSC (supported by U.S. Department of
Energy grant number DE-AC02-05CH11231),
and by the USQCD
collaboration.  This research used resources of the Oak Ridge Leadership 
Computing Facility at the Oak Ridge National Laboratory, which is supported 
by the Office of Science of the U.S. Department of Energy under Contract 
number DE-AC05-00OR22725. 
The PRACE Research Infrastructure resources  at the 
Tr\`es Grand Centre de Calcul and Barcelona Supercomputing Center were also used.
Parts of the calculations used the {\tt chroma} software
suite~\cite{Edwards:2004sx} and the {\tt quda} library \cite{Clark:2009wm,Babich:2011np}.  
EC was supported in part by the USQCD SciDAC project, the U.S. Department of Energy through 
grant number DE-SC00-10337,  and by U.S. Department of Energy grant number DE-FG02-00ER41132.
ZD, WD and PES were partly supported by  U.S. Department of Energy Early Career Research Award DE-SC0010495 and grant number DE-SC0011090.
The work of WD is supported in part by the U.S. Department of Energy, Office of Science, Office of Nuclear Physics, within the framework of the TMD Topical Collaboration.
KO was partially supported by the U.S. Department of Energy through grant
number DE-FG02-04ER41302 and through contract number DE-AC05-06OR23177
under which JSA operates the Thomas Jefferson National Accelerator Facility.  
MJS was supported  by DOE grant number~DE-FG02-00ER41132, and  in part by the USQCD SciDAC project, 
the U.S. Department of Energy through grant number DE-SC00-10337.	
BCT was supported in part by a joint City College of New York-RIKEN/Brookhaven Research Center
fellowship, and by the U.S. National Science Foundation, under grant
number PHY15-15738. 
MLW was supported  in part by DOE grant number~DE-FG02-00ER41132.
FW was partially supported through the USQCD Scientific Discovery through Advanced Computing (SciDAC) project 
funded by U.S. Department of Energy, Office of Science, Offices of Advanced Scientific Computing Research, 
Nuclear Physics and High Energy Physics and by the U.S. Department of Energy, Office of Science, Office of Nuclear Physics under contract DE-AC05-06OR23177.

\bibliography{bibi2.bib}
\end{document}